# DECOHERENCE IN A QUANTUM NEURAL NETWORK

Deniz Türkpençe*, Tahir Çetin Akıncı, Serhat Şeker

dturkpence@itu.edu.tr, akincitc@itu.edu.tr, sekers@itu.edu.tr

İstanbul Technical University, Department of Electrical Engineering,

34469, İstanbul, Turkey.

**Abstract**

In this study, we propose a spin-star model for spin-(1/2) particles in order to examine the coherence dynamics of a quantum neural network (QNN) unit. Since quantum computing paradigm promises advantages over their classical counterparts, quantum versions of neural networks can be examined in this context. We focus on quantum coherence as a natural resource for quantum computing and investigate the central spin coherence of a spin star model in the time domain in a dissipative environment. More particularly, we investigate the extent to which the central spin coherence time would be prolonged under specific parameters and spin-coupling Hamiltonians in a Markov environment. We find that Heisenberg XX-type couplings are more favourable for spin coherence time and the increase on the number of ambient spins extend the coherence time only in this coupling scheme. We also show that ising-type spin coupling is not desirable since it rapidly diminishes the coherence time in a dissipative environment.

## 1. Introduction

Parallel to the recent developments on artificial intelligence and machine learning, renewed interest to quantum versions to neural networks has been emerged. As the quantum computing and related technologies promises speed-ups over classical technologies (Nielsen and Chuang, 2011) it's fair to expect advantages from quantum neural networks over classical neuro-computing tasks (Hecht-Nielsen, 1990). Quantum coherence is an important resource for quantum computing, communication (Nielsen and Chuang, 2011) and quantum thermodynamics (Scully *et al.*, 2006; Türkpençe and Müstecaplıoğlu, 2016). For this reason, evaluating the quantum coherence of a spin system is of importance to the examination of the quantum version of a neural network model. The simplest neural network unit is a perceptron and an available representation of its quantum version would be a central spin system with anti-ferromagnetic interactions. While quantum coherence is a physical resource for quantum information processing, decoherence is an irreversible loss of valuable quantum information of quantum systems due to the inevitable reservoir interactions. A considerable effort has been performed for suppressing decoherence and extending the coherence time of quantum systems (Zhu W and Rabitz H, 2003; Korotkov AN and Keane K, 2010; Sadeghi SM and Patty KD, 2014) in order to exploit quantum resources for longer operation times. We propose a coupled central-spin system immersed in a Markovian bath in which the central spin has initial quantum coherence. We numerically show that central spin coherence time could be extended with respect to the increase of identical ambient spins with special coupling mechanisms over quantum interactions. The proposed model makes the scheme more feasible and less expensive for the goal of preserving quantum coherence by the use of natural coupling mechanisms instead of artificial and complicated decoupling mechanisms. The numerical simulation of the proposed model was based on the time evolution of the quantum master equations in Lindblad form. It has also been shown that quantum thermodynamics can exploit quantum coherence as a non-traditional resource for energy extraction from a single heat reservoir without violating the thermodynamic laws (Scully *et al.*, 2006; Türkpençe and Müstecaplıoğlu, 2016; Scully, 2010). However, quantum coherence is fragile and rapidly decays due to the coupling to the environmental degrees of freedom and dealing with quantum coherence for realistic systems requires an open quantum system approach (Breuer and Petruccione, 2007). Dynamical methods for decoherence suppression has been examined by various methods (Suter and Alvarez, 2016). However, suppressing decoherence by dynamical methods has high-energy costs and is not desirable for artificial energy harvesting systems depending on the coherence lifetime.

We adopt the Lindblad master equation formalism to simulate the open quantum system dynamics. In our study, we show that a single qubit coherence time can be extended in a Markovian environment by natural ferromagnetic interactions. We believe that our findings would be helpful for dynamical control methods for quantum systems dealing with neuro-computing tasks aiming to exploit the quantum resources.

## 2. Model and system dynamics

Central spin networks find various applications in quantum decoherence analysis (Zurek, 2006) and quantum communication (Chen *et al.*, 2006). In particular, we consider $N$ number of spin-

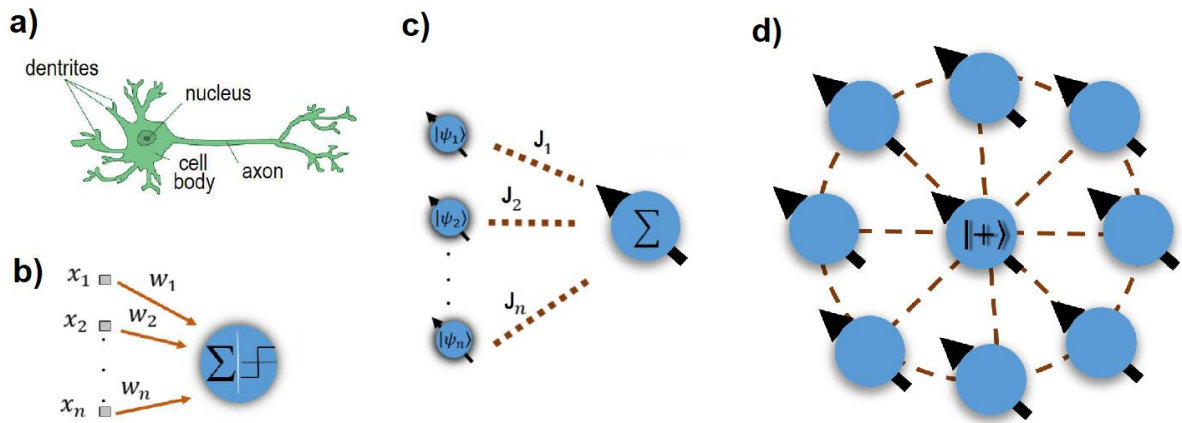

**Figure 1.** Biological, computational and quantum mechanical representations of simple neural nets. (a) A biological neuron (b) A perceptron with $N$ input nodes. (c) A quantum version of a perceptron (d) A central spin system with central initial coherence

1/2 particles (qubits) including the central one with quantum anti-ferromagnetic interactions as a central-spin network. Fig. 1 illustrates the central-spin network with ambient spins. As is clear in Fig.1 (a) and (b) a perceptron has similar structure with $N$ input nodes just like a biological neuron with dendritic connections. Neuro-computing tasks mostly based on binary McCulloch-Pitts neurons (McCulloch and Pitts, 1943) by this analogy. A quantum version of a perceptron (see Fig.1 (c)) could be considered as some number of input nodes interacting with a common output node. In the quantum version, these nodes are represented by spin-(1/2) particles (or quantum dots). In this model it's been recently shown (Türkpençe *et al.*, 2017) that in this model the output node experiences a coherent linear combination of the input states for closed system dynamics. On the other hand, for open quantum system dynamics, output node evolves into mixed combination of information reservoir states in which the input nodes are in contact with. During this evolution, the initial coherence of the output node vanishes.

The goal of this study is to analyse the output node coherence time of a central spin system (see Fig.1 (d)) which is in essence nothing but a quantum version of a perceptron model as in Fig.1 (c). In our study, we adopt the density matrix formalism in order to represent the system dynamics. We assume that initially all the individual qubits are in the tensor product form, and inter spin couplings are weak. The Hamiltonian describing the central-spin network reads

$$H = \sum_i^N \omega_i S_z^i + \sum_i^{N-1} J_\nu^i S_\nu^1 S_\nu^i \tag{1}$$

where $S_\nu$ are the spin operators with $\nu: x, y, z$. Here, $S_\nu^1$ is the central spin operator and $\omega_i$ is the corresponding Bohr frequency of the $i^{th}$ spin. Here, spin operators are

$$S_x = \frac{1}{2}\sigma_x, \ S_y = \frac{1}{2}\sigma_y, \ S_z = \frac{1}{2}\sigma_z \tag{2}$$

where $\sigma_x, \sigma_y$ and $\sigma_z$ are the Pauli operators. The units taken such as $\hbar = 1$ throughout the paper. We consider two initialization schemes. First, ambient spins obey a Boltzmann distribution very near to the ground state and in the second case ambient spins are initially in a cat state which is a superposition of orthogonal spin-up and spin-down states. We denote this state as $|+\rangle = \frac{1}{\sqrt{2}}(|\uparrow\rangle + |\downarrow\rangle)$. The initial state of the central spin is $|+\rangle$ in both cases. In the scenario, the system immersed in a Markovian bath and the central spin coherence time investigated under different bath and spin interaction parameters. Open quantum system dynamics governed by a Lindblad master equation (Breuer and Petruccione, 2007),

$$\dot{\rho} = -i[H, \rho] + \gamma \sum_i \mathcal{D}[\sigma_-^i] + \frac{\gamma_\phi}{2} \sum_i \mathcal{D}[\sigma_z^i] \tag{3}$$

where $H$ is the Hamiltonian, $\rho$ is the density matrix of the system and $\mathcal{D}[x] = (2x\rho x^\dagger - \{xx^\dagger, \rho\})/2$ is a Lindblad superoperator describing the relaxation to the environmental degrees of freedom. Here, $\{.\}$ represents anti-commutation, $\gamma$ and $\gamma_\phi$ are the atom decay rate and decoherence rate of each two-level system respectively. The atom decay and decoherence rates were assumed to be equal for each qubit for simplicity. In this study we consider the (J > 0) anti-ferromagnetic spin interactions for Ising and Heisenberg type Hamiltonians. As stressed above, the initial system density matrix is in the tensor product form of the individual density matrices as $\rho = \otimes_i^N \rho_i(0)$. Since the central qubit is the subsystem of interest, the reduced dynamics was obtained by using the partial trace operator over ambient spins degrees of freedom. In this respect, central spin dynamics obtained as $\rho_1(t) = \text{Tr}_i[\rho(t)]$. $l_1$ norm of coherence (Baumgratz et al., 2014) is a popular measure, which is nothing but the absolute sum of the off-diagonal elements of a density matrix. In our study, we use the observable $\langle \sigma_x \rangle = \text{Tr}[\rho \sigma_x]$ as the coherence measure alongside the $l_1$ norm of coherence. In the next section, we present the results of the evolution of the central spin coherence in the time domain by using the explained model, the coherence measures and the parameters of the central spin system.

## 3. Results

Environment induced decoherence is a common problem for quantum enabled technologies particularly aiming to benefit the quantum resources. For the goal of specific implementations, environment should be perfectly modelled and realistic parameters should be used for the current state of art.

However, in our study, we are interested in a more general situation with a generic Markovian environment as a zero temperature bath. Therefore, our results will be much more general than the particular system and baths. In this section we present the decoherence dynamics of the central spin of the proposed system depending on different coupling schemes. The effect of the number of ambient spins to the decoherence time was investigated in the numerical simulations. Thermal and maximally coherent (cat) states $|+\rangle$ were examined as two distinct cases for the initial states of the ambient spins. A thermal qubit state of a system with a Hamiltonian $H = \frac{\omega}{2}\sigma_z$ would be described as $\rho_{th} = \frac{1}{Z}e^{-\beta H}$ where $Z = \text{Tr}e^{-\beta H}$ is the partition function and $\beta$ is the thermal inverse temperature.

First, the Ising type spin Hamiltonian ($J_x = 0$, $J_y = 0$) is examined in order to investigate the central spin decoherence process as shown in Fig. 2 (a). Here, we observe that decoherence time explicitly reduces during the increase of thermal ambient spins interacting with the central spin with an Ising type spin Hamiltonian. Fig. 2 (b) also present the decoherence behaviour for the case of thermal ambient spins. In this case, the spin interactions are in the Heisenberg XXX-type ($J_x = J_y = J_z$). As is obvious in Fig. 2 (b), we report revival of coherence during the time evolution process. Revivals during equilibration are mostly referred to non-Markovian dynamics (Inés de Vega et *al.*, 2017). Here, we report this revival scheme of decoherence for a weakly interacting central spin system in a Markovian environment. A revival of coherence recently introduced experimentally for trapped atoms (Afek et *al.,* 2017).
 We observe that also in the Heisenberg XXX-type interaction case coherence time reduces with the increase of the number of ambient spins though the revivals are present. We remind that at the first initialization case the system is at the tensor product of the individual qubit states in which the ambient spins are in a thermal Gibbs state.

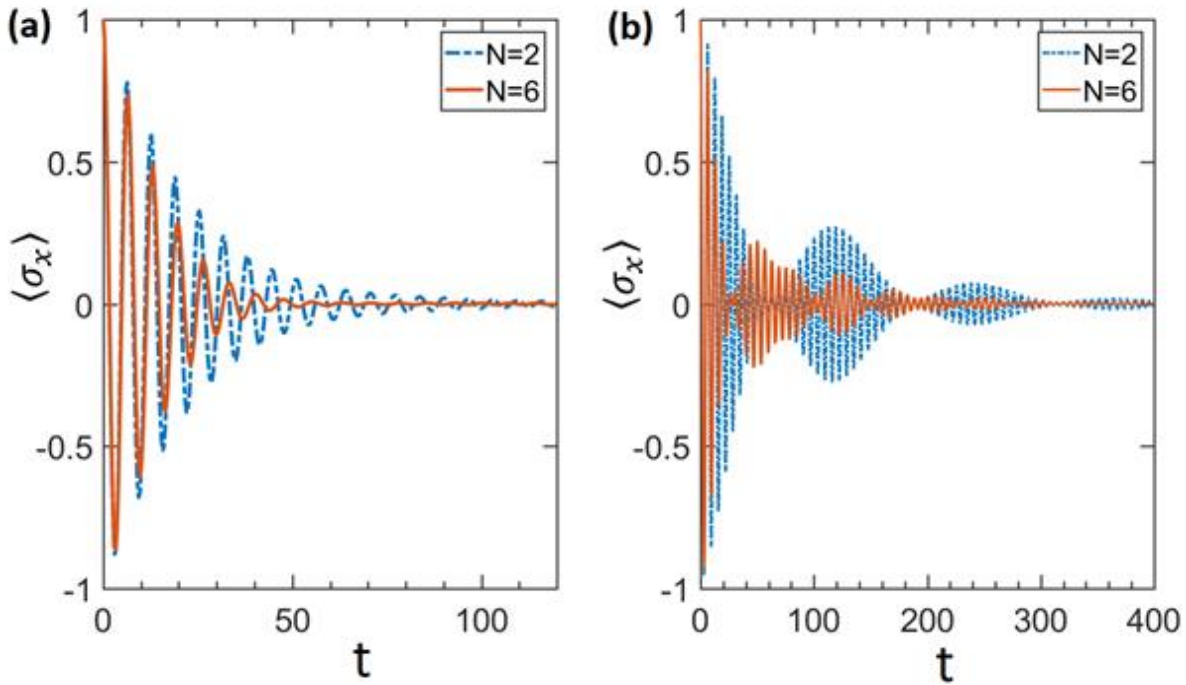

**Figure 2.** Time evolution of central spin coherence for thermal ambient spins. (a) Evolution of observable $\langle \sigma_x \rangle$ of the central spin in the presence of Ising type spin interactions. (b) Evolution of observable $\langle \sigma_x \rangle$ of the central spin in the presence of Heisenberg XXX-type spin interactions. Couplings between spins are taken $J = 0.05$. Inverse temperature $\beta = 0.5$ in units of $\hbar\omega/k_B$ for thermal ambient spins. Atom decay rate $\gamma$ and coherence decay rate $\gamma_\phi$ are equal and $\gamma = \gamma_\phi = 1.5 \times 10^{-2}$. $\gamma, \gamma_\phi$ and time are dimensionless and scaled by Bohr frequency $\omega$.

In this case, one can see that the increase at the number of thermal ambient spins enhance the reduction of central spin coherence time with an Ising type interaction Hamiltonian. Since the coherence time reduction also occurs in the presence of Heisenberg XXX-type Hamiltonian depending on the number of thermal ambient spins, we can conclude that thermal ambient spin states would be detrimental for central spin coherence. Though the revivals of coherence are interesting in this coupling scheme, one cannot trivially conclude that these revivals would be beneficial. Next, we investigate the case where all the ambient spins have coherent |+⟩ initial state. In this case central spin coherence behaves totally different than the initial case where all ambient spins are thermal. Again, we investigate the Heisenberg XXX-type spin interaction Hamiltonian as in Fig. 3 (a) and observe that decoherence time is independent of the ambient spin number since we obtain overlap plots depending on different $N$. Also the $l_1$ norm of coherence of this case has overlap plots as expected.

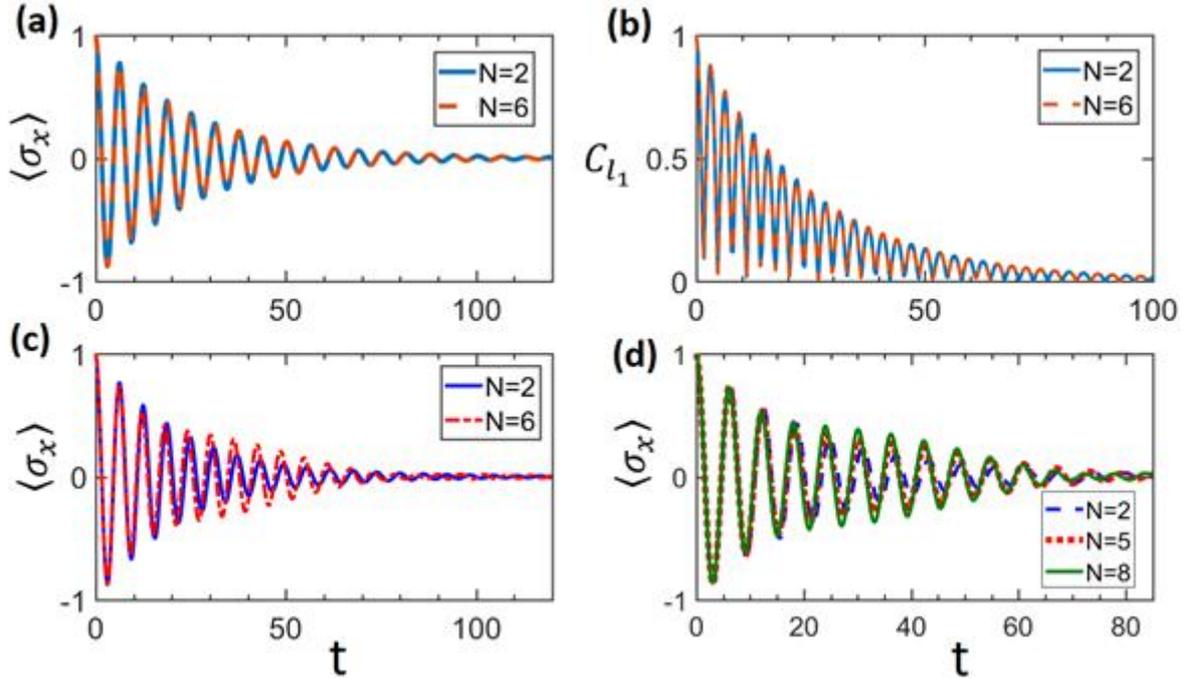

**Figure 3.** Time evolution of central spin coherence for coherent ambient spins. Time evolution of the observable (a) $\langle\sigma_x\rangle$ and the (b) $l_1$ norm of coherence of the central spin in the presence of Heisenberg XXX-type spin interactions depicted. Time evolution of the observable $\langle\sigma_x\rangle$ of the central spin in the presence of Heisenberg XX-type spin interactions presented for (c) $N = 2$, $N = 6$ and (d) $N = 2$, $N = 5$ and $N = 8$ number of ambient coherent spins. Couplings between spins are taken $J = 0.05$. Atom decay rate $\gamma$ and coherence decay rate $\gamma_\phi$ are equal and $\gamma = \gamma_\phi = 1.5 \times 10^{-2}$. $\gamma$, $\gamma_\phi$ and time are dimensionless and scaled by Bohr frequency $\omega$.

On the other hand, we observe that things change dramatically in the presence of Heisenberg XX-type interaction as in Figs. 3 (c) and (d). In Fig. 3 (c) it's obvious that there is an extension of coherence time in the presence of Heisenberg XX-type spin interactions with the initial $|+\rangle$ state. The comparison of Fig. 3 (a) and (c) reveals the effect of spin interaction type to the coherence time clearly. In this particular case, we observe that Heisenberg XX-type interaction causes an extension of central qubit coherence time during the increase of ambient spins. Comparison of Fig. 3 (a) and (c) also uncovers the fact that switching from Heisenberg XXX-type interaction to Heisenberg XX-type interaction gives rise to transition from ambient spin number independent coherence time to the ambient spin number dependent case. These results could be important for dynamic control of quantum neural systems. Note that for Heisenberg XX-type interaction, extension of central qubit coherence time occurs only in the presence of coherent ambient spins. Fig. 3 (d) supports this result by revealing the extension of central spin coherence time depending on the coherent ambient spin number up to 8 spins. Controlling the central spin coherence time by only natural spin interaction preferences instead of dynamical methods provides resource saving and prevents complicated interventions to the system which makes quantum control difficult. We believe that our results would contribute also to the quantum cognitive systems, based on bio-inspired neural systems for longer operation times.

## 4. Conclusion

We propose a central spin quantum system stands for a quantum version of a neural network unit with generic spin interaction Hamiltonians. We focus on the evolution of central spin coherence in the time domain under particular interaction Hamiltonians and ambient spin numbers. It's found that thermal and coherent ambient spins have different effects on the central spin coherence time. We also show that it's possible to reduce or extend the coherence time by an appropriate choice of spin interaction Hamiltonians depending on the ambient spin numbers. We numerically show that in the presence of Heisenberg XX-type spin interaction Hamiltonians central spin coherence time has been prolonged due to the increase of coherent ambient number of spins while Ising type spin Hamiltonians reveal an opposite behaviour. The presented results are of importance to the physical implementation of quantum computational protocols aspires longer operation times and to the quantum versions of cognitive systems in contact with environmental degrees of freedom. These results would also contribute the quantum control methods aiming the optimal benefits of quantum resources.

**Acknowledgements**

Authors acknowledge İstanbul Technical University for technical support. The authors declare that the research was conducted in the absence of any commercial or financial relationships that could be construed as a potential conflict of interest.